\documentstyle[12pt,preprint]{aastex}

\shorttitle{Halting Planet Migration}
\shortauthors{Kuchner \& Lecar}
\begin{document}
 
\title{Halting Planet Migration in the
Evacuated Centers of Protoplanetary Disks}
\author{Marc J. Kuchner\altaffilmark{1}}
\affil{Harvard-Smithsonian Center for Astrophysics \\ Mail Stop 20, 60 Garden St., Cambridge, MA 02138}
\altaffiltext{1}{Michelson Postdoctoral Fellow}
\email{mkuchner@cfa.harvard.edu}

\author{Myron Lecar}
\affil{Harvard-Smithsonian Center for Astrophysics, Mail Stop 51}
\email{mlecar@cfa.harvard.edu}

\begin{abstract}

Precise Doppler searches for extrasolar planets
find a surfeit of planets with orbital periods of 3--4 days,
and no planets with orbital periods less than 3 days.
The circumstellar distance, $R_0$, where
small grains in a protoplanetary disk reach sublimation
temperature ($\sim 1500$~K) corresponds to a period of
$\sim 6$ days.  Interior to $R_0$, turbulent accretion due
to magneto-rotational instability may evacuate the disk
center.  We suggest that planets with orbital
periods of 3--4 days are so common because migrating planets halt
once this evacuated region contains the sites of their
exterior 2:1 Lindblad resonances.

\end{abstract}
 
 
\section{Introduction}

Analytic calculations \citep{gold80,ward97a} and
numerical simulations \citep{nels00, kley01} suggest that
protoplanets in a protoplanetary disk migrate
rapidly into the star they orbit---so rapidly that it is
a wonder any planets survive at all.
Small protoplanets torque the disk at
Lindblad and corotation resonances, and the resulting
back-torque can propel a planet into the
star in a matter of $10^{5}(M_P/M_{\bigoplus})^{-1}$
years \citep{ward97b}.
\citet{ward97b} has dubbed this conundrum the Shiva
problem, after the Hindu god of destruction.
Large protoplanets may open a gap in the disk via their
resonant torques, and so
become locked to the disk's viscous spreading,
a process which may dump the most of the disk and planet onto the
star within $10^{7}$ years or less, depending on the disk
viscosity.

\begin{figure}
\epsscale{1.0}
\plotone{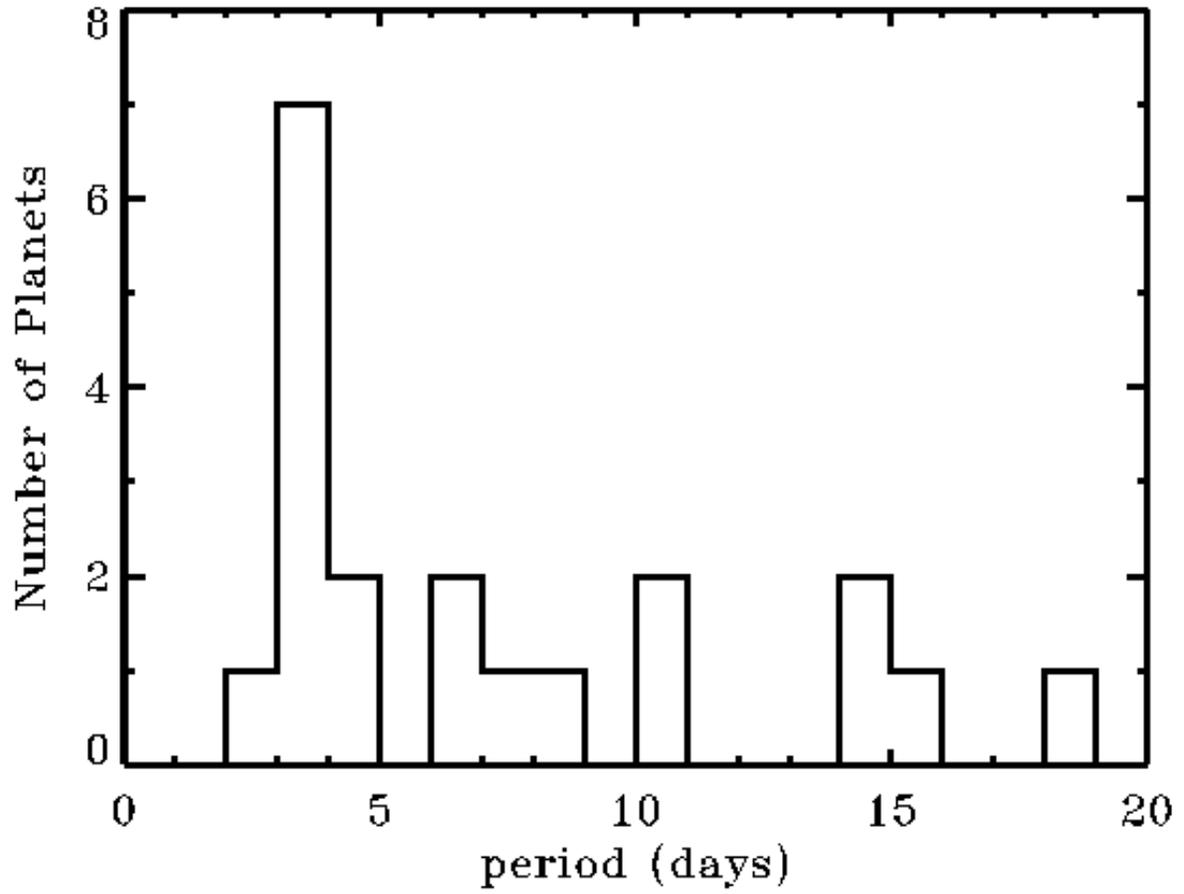}
\caption{Histogram of the orbital periods of the extrasolar
planet candidates detected by the Precision-Doppler technique.}
\label{fig:ahistogram}
\end{figure}

Figure~\ref{fig:ahistogram} shows the distribution
of the orbital periods of the innermost Doppler planet
candidates, summarizing data from
\citet{exoplanets.org}\footnote{This website about
extrasolar planets is available at 
\url{http://exoplanets.org/}}.  These candidate planets all
orbit stars with masses in the range 0.7--1.4 $M_{\odot}$.
Figure~\ref{fig:ahistogram} suggests that whatever mechanism
halted the migration of these planets operates best
at an orbital period of $\sim 3$~days and
ceases to operate at shorter periods.  Of the 20 planets with
periods less than 20 days, 8 have periods in the range 3--4 days.
No planet has a period less than 2.98 days.
This trend appears to be real and not an artifact of observational
selection;  the primary precise-doppler surveys are complete for
Jupiter mass planets out to a period of $\gtrsim 0.5$
years \citep{butl01}.

Occasionally the interactions
among two planets and a star can leave a planet
trapped by stellar tides into a circular orbit at $\sim 0.04$ AU
\citep{rasi96}.  Sometimes an accreting planet may overflow its
Roch lobe, losing mass to the star, and this process may
halt the planet's migration \citep{tril98}.
But these phenomena appear to be rare.

Scenarios in which planets migrate by
interacting with a disk of planetesimals without gas
\citep{murr98} provide a disk truncation
radius where planets may gather: the radius where
planetesimals become hot enough to sublimate.
The problem with gas-less migration schemes is that
substantially changing the orbit of a Jupiter-mass
planet requires roughly a Jupiter mass of
planetesimals.

In contrast, optically-thick disks with
more than a Jupiter mass of gas appear to be ubiquitous
around young stars. \citet{lin96} have suggested that planet
migration ends where the gaseous protoplanetary
disk meets the stellar magnetosphere.  We suggest an alternative explanation for
the pile-up of planets near the 3-day period: a gas disk truncated
at a temperature of 1500~K by the onset of
Magneto-Rotational instability
(MRI; Chandrasekhar 1961; Balbus \& Hawley 1991).

\section{The Theory}

A planet with orbital period $p$ and semimajor axis $a$
interacts with a disk at Lindblad resonance
sites, located where the natural epicycle period of
the gas in the disk (roughly the Keplerian orbital period) is
$p(m+1)/m$.
The most distant Lindblad resonances from the planet are
the $m=1$ and $m=-2$, located at roughly $1.59 a$ and $0.63 a$
respectively.  The planet torques the disk at these resonances,
and the back-torque from
the inner Lindblad resonances tends to add angular momentum
to the planet's orbit, while the back-torque from the 
outer Lindblad resonances tends to remove angular momentum to the
planet's orbit \citep{gold80}.  However, over a wide range of
disk pressure and density gradients, the torques from the
outer Lindblad resonances dominate the torques from the inner
Lindblad resonances, 
causing the planet to spiral into the star \citep{ward97a}.

Consider a planet contained in a protoplanetary disk with
an inner truncation radius of $R_0$, corresponding to an orbital
period of $P_0$.  When the planet migrates inward to
where it has an orbital period $p \lesssim 2 P_0$
($a \lesssim 1.59R_0$),
it may begin to migrate inwards faster than normal because its
interior Lindblad resonance sites lie in the evacuated
regions, so the outward torques from these
resonances disappear.  When the planet reaches an
orbital period of $1/2 P_0$ ($a = 0.63 R_0$), and all of
the planet's inner and outer Lindblad resonances are contained
in a central evacuated region, the disk-planet interaction may
stall, ending the planet's migration \citep{lin96}. 
A planet with a gap is not immune; it must also stop migrating
at a period of $1/2 P_0$, where it can no longer dynamically
communicate with enough mass in disk material to substantially affect
its orbit.

We propose that protoplanetary disks
are evacuated interior to radius $R_0$ because of
the powerful magneto-rotational instability
which afflicts conducting
Keplerian disks.  Protoplanetary disks may
suddenly become conducting where they reach a temperature
of 1500K and potassium ionizes \citep{ston00}
and dust sublimates, removing recomination sites \citep{sano00}.
Numerical simulations of saturated MRI show that it can
provide a disk $\alpha$ viscosity \citep{lynd74} in the 
range of $\alpha \approx 0.004$ to $\alpha \approx 0.1$
\citep{hawl95, zieg01} depending on the assumed resistivity.
Exterior to $R_0$, the MRI may be damped completely
\citep{gamm96, reye01}.

In a steady state disk, the accretion rate
$\dot M$ is uniform and continuity dictates that 
$R \Sigma  v_R=$ constant, where $R$ is the
radial coordinate, $\Sigma$ is the surface density, and
$v_R$ is the radial drift velocity.
In a viscous accretion disk, $v_R \sim \nu/R$,
where $\nu$ is the viscosity.  So roughly speaking,
$\Sigma \propto \nu^{-1}$, or using the $\alpha$
prescription for disk viscosity, $\Sigma \propto \alpha^{-1}$.

Magnetorotational instability can plausibly cause  
the effective $\alpha$ to climb by a factor of 10--1000
interior to $R_0$, which translates into a comparable drop in
surface density.   When planets reach $a = 0.63 R_0$ in
the centers of disks evacuated by MRI-driven accretion,
their migration will slow by this factor of 10--1000.
The high viscosity in this zone also acts to inhibit gap
formation and close up existing gaps around massive planets.

In radiative-transfer models of
protoplanetary disks, a surface layer of small
grains intercepts the starlight first and rations it
to the disk midplane \citep{calv91, malb91, chia97, sass00}.
These small grains have hotter equilibrium temperatures
than a blackbody at a given distance from the star,
so they sublimate farther from the star than larger grains.
We can find $R_0$ by iteratively
solving the radiative equilibrium equation for the temperature
of the small grains as a function of distance from the star \citep{back93}.
We performed such a computation assuming $\sim 0.1 \mu$m
grains---bodies whose emissive and absorptive efficiencies
stay constant at wavelengths shorter
than 0.1~$\mu$m, but decline as $\lambda^{-1}$ at
longer wavelengths.  We find that the grains reach 1500~K
at $R_0=0.067$ AU ($P_0=6.3$ days) for a solar type star,
or $R_0=0.055$ AU ($P_0=4.8$ days) for star with solar luminosity
and mass, but effective temperature of $4000$~K, a
common model for a T Tauri star. 

Planet migration in a disk with such an inner truncation radius 
would halt at $p \approx P_0/2 =$~2.4~days; near the
observed pile-up of precise-Doppler planets.
The luminosity of a solar-mass
star varies over roughly a factor of $\sim 5$ during
the time from 1 million to 10 million years after
its birth \citep{dant94}, so the value of $P_0$ 
relevant to halting the migration of a given planet
could conceivably range over a factor of $\sim 2$
due to this effect.  For comparison, if we assume
a 1 Jupiter mass planet has a radius of 2.3 Jupiter
radii at an age of 1 million years \citep{guil02}, we
calculate that the orbital period where it overflows
its Roche radius is $\sim 1.3$~days.

\section{Discussion}

\subsection{OBSERVATIONS OF PROTOPLANETARY DISKS}

The likely hosts of protoplanetary disks are
T Tauri stars ($ \lesssim 2 M_{\odot}$) and
Herbig Ae/Be stars ($\sim$ 2--8$ M_{\odot}$).
Near-infrared photometry of Herbig Ae/Be stars
reveals a curious pattern; many of these stars
have two-humped infrared excesses--broad emission peaks
in the mid- or far-infrared, and then second
emission peaks near $3 \mu$m \citep{lada92}.
A model in which accretion luminosity
powers the near-IR hump requires optically-thick gas interior to
the dust-sublimation point, a feature not corroborated
by the observed SEDs \citep{hart93} or measurements of optical
veiling \citep{ghan94}.  The near-IR humps appear to
represent the thermal emission peak of a population of dust
at a temperature of 1500 K \citep{sylv96, malf98}.
The prominence of these near-IR humps may indicate
the importance of $R_0$ in the structure of Herbig Ae disks.

Further evidence for the structural importance of $R_0$ comes from
long-baseline optical interferometry.
\citet{mill01} resolved 11 Herbig Ae stars at baselines
of 21 and 38 m with the IOTA interferometer in the near-infrared.  
The visibilities of the 11 resolved targets and
their near-IR excesses were well modeled by dusty rings 
with inner radii of 0.3--5 AU, possibly
corresponding to $R_0$.  Measurements of the near-IR sizes of the
central regions of other young stars, including some T Tauri stars,
with the Palomar Testbed Interferometer \citep{akes00} and 
using a single Keck telescope with an aperture mask
\citep{tuth01, danc01} support this association; the observed
near-IR sizes show a trend of increasing with the
luminosity of the young star, which is consistent with the
increase of $R_0$ with luminosity
(J. Monnier \& R. Millan Gabet, in preparation).

\citet{natt01} and \citep{dull01} have interpreted the
photometric and interferometric observations of Herbig Ae/Be stars
as signs that protoplanetary disks may have tall,
passively-heated inner walls; hot dust on the
wall provides the $3 \mu$m emission
peak.  This inner wall could conceivably
represent the zone where the opacity and the surface density
suddenly drop in our model.  
Further high-resolution interferometric studies of
Herbig Ae/Be and T Tauri disks should help clarify how disk
structure affects planet migration.

\subsection{MAGNETIC DISK TRUNCATION AND THE X-WIND THEORY}

\citet{shu00} review a theory of the
interiors of protoplanetary disks which centers on
the interaction between the stellar magnetic field and
the inner edge of the conducting gas disk.
In this ``X-wind'' theory, the disk is replaced by
magnetic accretion columns and a bipolar wind interior
to the radius $R_x$.  Earlier explanations of
halting planet migration \citep{lin96} have
pointed to $R_x$ as the critical radius rather
than $R_0$.

Directly measuring $R_x$ is difficult.  However,
in equilibrium, the stellar rotation period matches the
Keplerian period at $R_x$,
so measuring the stellar rotation rate and assuming
that T Tauri stars are rotationally
locked to their disks may provide an estimate of $R_x$.
However, the process of breaking the stellar rotation
is probably slow \citep{hart02}, and the rotation periods
of T Tauri stars range from
0.5 to $\ge 8$ days \citep{stas99}.  Consequently,
it is hard to associate the 3-day
period of the closest-in extrasolar planets with a
special stellar rotation rate, or with $R_x$.

In the X-wind model of chondrule formation, 
a geometrically-thin, optically-thick disk of
planetesimals interior to $R_x$ suffers mutual
collisions and heating from magnetic flares
\citep{shu96, shu01}.  Material from these
planetesimals become proto-chondrules.
In our picture, Type I migration
helps deliver planetesimals to the magnetically-active
zone.  We imagine a ring of planetesimals collecting
near $0.63 R_0$.  Objects smaller than a few km
may migrate in further than $0.63 R_0$ via plasma drag
\citep{shu97,shu01}.

\subsection{TORQUE REVERSAL}

We briefly considered the following alternative mechanism
for halting planet migration.
Continuing the calculations of \citet{ward97a}
shows that with only a slight positive temperature gradient
($(R/T)dT/dR \gtrsim 0.6$, i.e. the disk temperature
increases with radius faster than roughly $R^{0.6}$),
the torque on the planet from disk tides can reverse.
Such a positive temperature gradient could conceivably
occur where the disk regains
sight of the star beyond the shadow of the disk's tall
inner wall.  However, the models of
\citet{dull01} suggest that the positive temperature
gradient in this region is largely confined 
to the surface layer of the disk, which does not have
enough surface density itself to affect planet migration.

\section{Conclusion}

We have painted a picture of the central regions of
protoplanetary disks in which a low-surface-density,
rapidly accreting central region catches
inwardly-migrating planets like flypaper.
A good way to test our conjecture that disk
temperature---not the stellar magnetic field---determines
the orbital radii of the innermost surviving
planets is to extend the sample of planet search
target stars to include a wider range of stellar
masses. For example, we predict that planets
around early A main sequence stars will collect at
a radius much farther from the star ($\sim 0.3$ AU)
than planets around solar type stars.   The high rotation
rates of main sequence A stars betray that
magnetic disk truncation can not rescue 
planets migrating into these stars until they reach
near 0.03 AU.  Doppler planet search
techniques may not work for A stars
because of their rotationally-broadened spectral lines.
However, astrometric searches, transit
searches, and direct-imaging searches for extrasolar
planets are not so limited.

\acknowledgements

We thank Joe Hahn, Doug Lin, Dimitar Sasselov, Hsien Shang and
Bill Ward for helpful conversations.

This work was performed in part under contract with the Jet Propulsion 
Laboratory (JPL) through the Michelson Fellowship program funded by 
NASA as an element of the Planet Finder Program.  JPL is managed for 
NASA by the California Institute of Technology

\end{document}